# Optothermally Controlled Charge Transfer Plasmons in Au-Ge$_2$Sb$_2$Te$_5$ Core-Shell Assemblies

*Burak Gerislioglu\*, Arash Ahmadivand, and Nezih Pala*

*Department of Electrical and Computer Engineering, Florida International University, 10555 W Flagler St., Miami, FL 33174, USA*

\*Corresponding Author: *bgeri002@fiu.edu*

**Abstract-** Tunable plasmonic resonances across the visible and near infrared spectra have provided novel ways to develop next-generation nanophotonic devices. In this study, by using optothermally controllable phase-changing material (PCM), we successfully induced highly tunable charge transfer plasmon (CTP) resonance modes. To this end, we have designed a two-member dimer assembly including gold cores and Ge$_2$Sb$_2$Te$_5$ (GST) shells in distant, touching, and overlapping conditions. We successfully demonstrated that toggling between amorphous (dielectric) and crystalline (conductive) phases of GST allows for achieving tunable dipolar and CTP resonances along the near-infrared spectrum. The proposed dimer structures can help forming optothermally controlled devices without further geometrical variations in the geometry of the design, and having strong potential for advanced plasmon modulation and fast data routing.

**KEYWORDS:** *Charge transfer plasmons, metallodielectric dimer, phase-change material, switching, plasmonics.*

**Introduction**

Effective control over the plasmonic characteristics of subwavelength metallic platforms is a significant approach for developing next-generation nanophotonic devices.[1,2] Such a capability over the spectral properties of nanoplasmonic structures, which are designed based on conventional metals (i.e. Au, Al), is difficult due to inevitable limitations in manipulating the dielectric functions of these solids.[3] Lately, various methods have been proposed to improve the tunability of optical properties of plasmonic structures, such as electrically controllable 2D materials (i.e. $MoS_2$, graphene)-based optoelectronic devices,[4-8] and electrochemical methods (i.e. electrochemical charging of particles).[9] Nevertheless, in most of these techniques, obtaining simultaneously reversible, tunable, and large plasmon resonance shift is still challenging. As another practical strategy, researchers have provided light intensity dependent tunable spectral responses by reducing the gap distance between nanostructures down to atomic scales.[10] According to the quantum mechanical theory for plasmonics,[11] shuttling of charges across the quantum gap *via* tunnelling gives rise to formation of a new type of resonance at energy levels apart from classical multipoles, known as charge transfer plasmon (CTP).[12-15] Even though this technique results in the formation of new plasmonic modes, it has limited spectral tunability.

Recently, it is shown that using optothermally controllable components (phase-change materials (PCMs), i.e. vanadium dioxide ($VO_2$),[16,17] AgInSbTe,[18] and $Ge_2Sb_2Te_5$ (GST)[19-22]) in the design of plasmonic structures helps to modify the corresponding spectral response without requiring any geometrical variations. In Ahmadivand *et al.*,[12] we demonstrated that bridging two neighboring nanodisks around a GST nanowire allows us to efficiently manipulate the shuttling of charges across an interconnecting junction by controlling the phase condition of the GST bridge. GST can maintain its state without requiring continuous energy, because of its non-volatility property.[19,23] Typically, crystallization process (transition from dielectric to conductive state) is achieved by annealing the material up to critical temperature around ~750 K,[24] which is primarily depends on the total volume of the PCM structure and the crystallization duration. The reverse transition (amorphization) can be obtained by melting and quenching the material over ~873 K in a short duration (~ns).[19] To this end, Joule heating, convective heating or ultra-fast optical stimuli can be used for both annealing and melt-quenching.

In this report, by using a pair of metallodielectric nanodisks consisting of gold core and GST shell, we analyze the excitation of optothermally steerable CTPs. We show that a two-member nanodimer system in three different regimes can be tailored to support controllable CTPs across the near-infrared region (NIR), when the GST shell is crystallized. Conversely, in the amorphous regime, the CTP peak disappears and classical dipole mode, due to capacitive coupling becomes dominant. To analyze this spectral behavior, we placed the core-shell type nanoparticles in not-touching, touching, and overlapping regimes and compared the results. Using full numerical analyses based on finite-element (FEM) and finite-difference time-domain (FDTD) methods, we studied the behavior of each orientation of core-shell nanodisks to achieve highly tunable CTP resonances.

## Results & Discussions

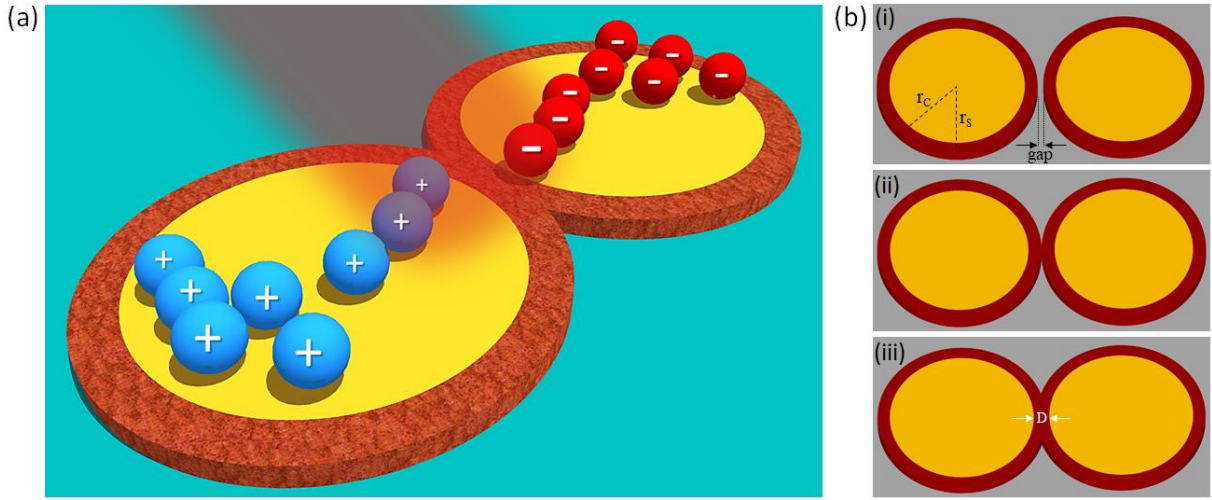

**Figure 1. (a)** An artistic rendering image of the metallodielectric core-shell dimer platform. **(b)** The proposed system in three different alignments with the corresponding geometrical parameters inside: **(i)** not-touching regime, **(ii)** touching regime, **(iii)** overlapping regime.

Figure 1a shows an artistic picture of the proposed metallodielectric nanodimer, and Fig. 1b(i-iii) indicates the platform in three different orientations with geometrical parameters description inside (not to scale). The substrate for all dimers is glass (SiO$_2$) with the relative permittivity of ~2.1, which is obtained from Palik's measurements.[25] The dimer core is gold (Au) with the dielectric function taken from empirically measured Johnson-Christy constants.[26] The Au disks are walled with GST shells, and the corresponding complex permittivity for both amorphous (a-GST) and crystalline (c-GST) phases were acquired from the experimental measurements.[27] In the electromagnetic analyses, we assumed that the metallic core is in perfect contact with the surrounding PCM shell layer, and the thickness of the dimer is fixed to 60 nm homogenously for all the investigated structures. The numerical modelling and switching between opposite phases of GST are explained in the Methods section.

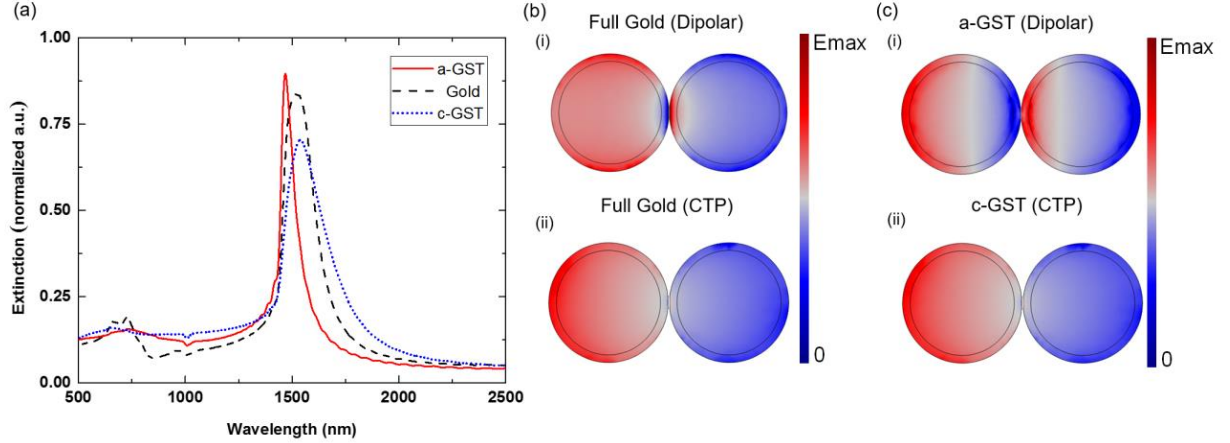

**Figure 2. (a)** Normalized extinction spectra of the nanodimer platform with full metallic, a- and c-GST shells. For the full metallic dimer, the extinction spectrum shows a dipolar peak at 700 nm, and a CTP resonance at 1525 nm. For the c-GST case, the CTP mode is appeared at 1550 nm, and for the a-GST dimer, the induced mode turns into dipolar resonance and blue-shifted to 1470 nm. Charge distributions at the extinction resonances for the nanostructure in Fig. 1a: **(b) (i)** full metallic (dipolar), **(ii)** CTP, **(c) (i)** a-GST (dipolar), **(ii)** c-GST (CTP). The E-field is set along the dimer axis.

Next, we investigate the dimer in fully metallic condition with the radius of 117 nm, and edge-to-edge separation of 0.5 nm. In this limit, the excited charges can shuttle across the gap, resulting with the accumulation of opposite charges in each side of system.[28] This effect can be recognized by the formation of a CTP peak (Fig. 2a (dashed curve)), which is consistent with previous quantum plasmonic studies for plasmonic nanostructures.[10] Continuing the study with metallodielectric core-shell dimer with a quantum gap of 0.5 nm, we examined the spectral response of the nanoparticles pair with the following geometries $(r_c, r_s)$=(102, 117) nm. In the amorphous state, the GST acts as a dielectric substance, and capacitive coupling becomes dominant.[29] Considering the overall width of the GST shell in both sides of dimer as $r_s-r_c$=15 nm, the interparticle distance between dimer components is 30 nm, hence, according to plasmon hybridization theory for plasmonic assemblies,[29] a significant dipole moment is arisen around 1495 nm (Fig. 2a (solid curve)). Besides, a weak quadrupolar shoulder is also appeared in the shorter wavelengths (~750 nm). Switching the phase of the GST shell to the crystalline state allows direct shuttling of the charges across the atomic opening, and the resulting spectral response (dotted curve in Fig. 2a) is analogous to the full metallic regime, as expected. We also observed a classical dipolar mode in the shorter wavelengths, owing to the metallic cores coupling.[30] In Fig. 2b, the charge distributions at the extinction resonances for the full Au dimer at both dipolar and CTP peaks positions are illustrated. As shown in Fig. 2b(i), the excited bright dipolar modes from Au cores capacitively coupled to each other, while for the CTP mode, the charges are split between the nanodisks (see Fig. 2b(ii)). In the presence of the GST shell (Figs. 2c(i, ii)), capacitive coupling and charge transfer events for a-GST and c-GST are become dominant, respectively.

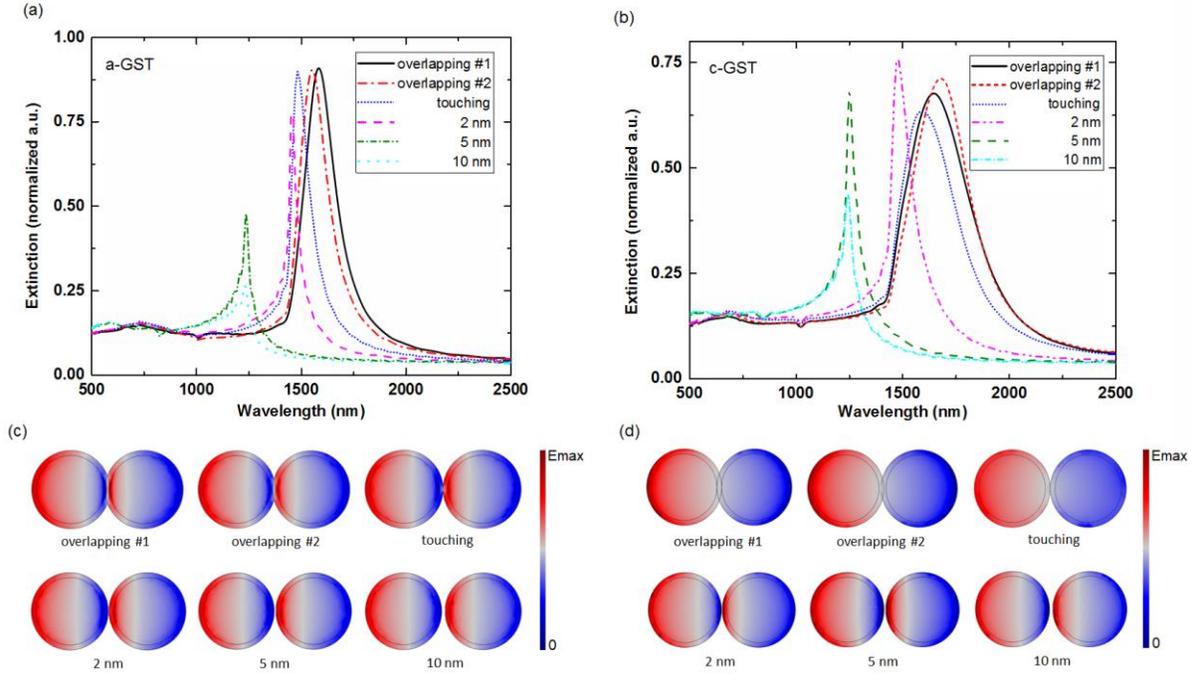

**Figure 3. (a), (b)** Normalized extinction spectra of the a-GST and c-GST dimer system, while the gap is variant (overlapping #1 and #2, touching, 2 nm, 5 nm and 10 nm), respectively. **(c), (d)** The corresponding charge distributions for different gap variations for a-GST and c-GST cases, respectively. (CTP). The E-field polarization is set along the dimer axis.

As mentioned above, the phase of the GST shell plays a substantial role in determining the quality, position, and behavior of CTP and dipole modes. In addition, the geometry of the established conductive region has an undeniable impact on the transfer and accumulation of the excited charges.[30] Both conductivity and resistance of the pathway can affect the charge transport. In the current structure, the touching junction width ($\delta_x$) is defined by: $\delta_x = 4r_2 \exp(-\pi/2 \times tR\sigma(\omega))$,[31] where $R$ is the entire resistance, $t$ is the thickness of the dimer, and $\sigma(\omega)$ is the frequency-dependent conductivity of GST. Accordingly, increasing the touching or overlapping cross-area (conductive regime) allows for the transfer of more charges and excitation of distinct CTP resonances. For the overlapping regime, we used "$D$" to show the overlap depth, which is $D=0$ nm for the touching case. Moreover, Fig. 3a exhibits the normalized extinction cross-section for the dimer assembly with a-GST shell, while the gap is varying gradually from capacitive coupling to touching and overlapping regimes. At first, we observed dipolar peaks around 1200 nm with capacitive gap from 10 nm and 5 nm, while the quadrupolar shoulder is dampened drastically (Fig. 3a). For 2 nm opening, the dipolar peak is red-shifted to ~1500 nm and intensified strongly. For the touching dimer, the dipolar peak enhances, due to reduced gap spot between the gold cores. Overlapping the nanoparticles with two different depths, specified by overlapped #1 ($D=15$ nm) and overlapped #2 ($D=22$ nm), allows to further enhancement in the dipolar peak and shift the position to the longer wavelengths. Moreover, for the dimer with c-GST shell and distant nanoparticles, the capacitive coupling is still dominant, owing to the absence of either quantum tunneling or direct charge transfer (Fig. 3b). In the touching and overlapping regimes, we facilitated direct shuttle of charges across the dimer structure. By increasing the cross-section of pathway, the CTP extreme red-shifts to the longer spectra with higher intensity. Figures 3c and 3d are the local charge density maps for the studied several cases for the metallodielectric dimer, consistent with the previous analyses for both dipolar and CTP modes.

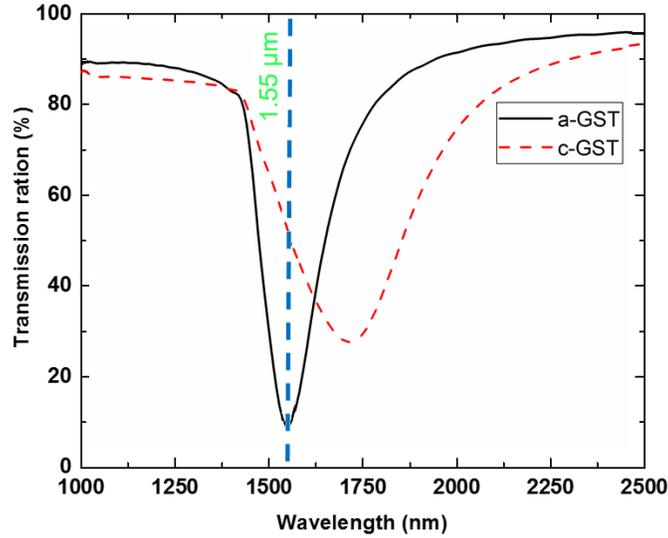

**Figure 4.** The transmission ratio of the GST based core-shell switch in a-GST and c-GST states.

Next, we examine the potential application of the proposed structure to be employed for advanced photonic switching purposes. Using the optothermally controllable metallodielectric structure, we tailor an active and efficient switch based on switching between dipolar and CTP resonances with the modulation depth up to 49% (Fig. 4). For the a-GST shell, a pronounced transmission dip appears at 1550 nm (global telecommunication wavelength), resembling ON state of the proposed switch. On the contrary, by switching the shell layer to c-GST, the dipolar dip disappears and a CTP mode rises around ~1750 nm, resembling the OFF state of the switch. Such ~200 nm shift in the position of resonant dip allows for optical signal modulations in practical applications. The achieved results are comparable with advanced all-optical Fano switches, and analogous PCM-mediated modulators based on complex designs and metamaterials.

**Conclusions**

In conclusion, a plasmonic dimer consisting of a pair of metallodielectric core-shell nanoparticles was analyzed for its optothermally controllable plasmonic resonances in the NIR. Using the simplest member of nanoparticles clusters family, we have studied the excitation of tunneling and direct CTP resonances by numerical tools. Choosing three different orientations of nanoparticles (distant, touching, and overlapping), and taking the advantage of thermally tunable GST shell, we demonstrated an efficient interplay between pronounced dipolar (in amorphous regime) and CTP (in crystalline regime) resonances. The obtained results constitute for the original observation of reversible CTP *via* either quantum tunneling or direct charge transfer in GST-mediated metallodielectric dimers. Exploiting the interplay between dipolar and CTP modes in opposite phases of the GST shell, we developed a platform for switching purposes at the telecommunication band with reasonable modulation depth.

**Methods**

For the numerical analyses, we used Lumerical FDTD and COMSOL Multiphysics. The incident beam was a broadband plane wave with the bandwidth of 400 nm-1600 nm, with the irradiation power of $P_0$=3.2 μW, beam fluence of 60 Jm$^{-2}$, pulse duration of 500 fs, and repetition of 10 KHz. The boundaries were surrounded by highly absorptive perfectly matched layers (PMLs). We also used additional light source with the duration of 0.9 ns and irradiation power of 5.5 mW to provide the required energy for the amorphization process. Besides, mesh with minimum lateral size 0.2 nm was applied, and the simulation time step is set to $dt$~0.1 fs based on the Courant stability.[32] For the optical analysis of the GST nanoshells, we used the effective permittivity at crystallization level based on Lorentz-Lorenz effective-medium expression:[33]

$$\left(\frac{\varepsilon_{eff}(\lambda)-1}{\varepsilon_{eff}(\lambda)+2}\right) = f_c\left(\frac{\varepsilon_c(\lambda)-1}{\varepsilon_c(\lambda)+2}\right) + f_a\left(\frac{\varepsilon_a(\lambda)-1}{\varepsilon_a(\lambda)+2}\right) \quad (1)$$

Where $f_i$ is the volume function of the *i*th phase as: $0 \le f_i = n_i/\sum_j n_j \le 1$, in which $n_j$ is the corresponding density of the *j*th phase.

**Acknowledgements**

This work is supported by Army Research Laboratory (ARL) Multiscale Multidisciplinary Modeling of Electronic Materials (MSME) Collaborative Research Alliance (CRA) (Grant No. W911NF-12-2-0023, Program Manager: Dr. Meredith L. Reed).